\documentclass[aps,pre,twocolumn,superscriptaddress,floatfix]{revtex4-2}

\usepackage{amsmath,amssymb}
\usepackage{graphicx}
\usepackage{booktabs}
\usepackage{bm}
\usepackage[colorlinks=true,allcolors=blue]{hyperref}

\newcommand{\Ut}{U_{3}}
\newcommand{\Up}{U_{2}}
\newcommand{\eps}{\epsilon}

\begin{document}

\title{Trimers, pairs and localisation on disordered random graphs}
\author{Alexei Vazquez}
\email{alexei@nodeslinks.com}
\affiliation{Nodes \& Links Ltd, Salisbury House, Station Road, Cambridge, CB1 2LA, UK}
\date{\today}

\begin{abstract}
Higher-order networks take group interactions as given. I ask what
happens when a group interaction is made by pairwise dynamics on a
graph, and follow it through the transition that decides whether it
matters at large scale, Anderson localisation. An emergent hyperedge is
a $k$-body bound state of a Hamiltonian whose terms live on the vertices
and edges of the graph, Borromean when none of its $(k-1)$-body faces is
bound, detected on a finite graph by a $k$-fold coincidence that
survives the thermodynamic limit while the $(k-1)$-fold ones vanish.
Sparse random graphs produce one from the simplest pairwise dynamics:
three distinguishable particles with on-site attraction $U$ and hopping
$t$ bind into a trimer before any pair binds. The thresholds are
obtained on the Bethe lattice, in closed form for the pair, in both
sectors that an expander graph has and a lattice lacks, the one a
condensate sees and the one a Fermi liquid sees. On a disordered graph
the emergent hyperedges nucleate on rare deep sites and localise before
their parts: the compact trimer, with a narrow band and a threefold
sensitivity to site energies, undergoes the Anderson transition at a
disorder of $1.6t$, against $17t$ for a single particle and $4t$ for a
pair, and the Fermi liquid of trimers is a glass above a line between
$1.5t$ and $1.7t$, where a liquid of single particles at the same density
needs $15t$ to $16t$. The transition is the single-particle cavity
transition with the composite's hopping and effective disorder. A group
interaction made by pairwise dynamics is an order of magnitude more
fragile to heterogeneity than the network that makes it.
\end{abstract}

\maketitle

\section{Introduction}
Network science has spent the last decade adding group interactions to
graphs. Hypergraphs and simplicial complexes represent interactions
among three or more units at once, and on them contagion,
synchronisation, consensus and percolation behave differently from their
pairwise versions~\cite{battiston2020,battiston2021,bick2023,bianconi2021,lambiotte2019,iacopini2019,neuhauser2020,sun2023}.
In all of this the hyperedge is an input: measured, as a co-authorship
or a protein complex, or postulated. The converse question is asked less
often. Given a graph whose edges carry only pairwise couplings, can the
dynamics on it produce an object that is irreducibly three-body, a group
that exists although no pair within it does? If it can, then a group interaction need not be a datum of the
system: it may be what pairwise dynamics on the underlying graph makes,
carried by a property of the group that none of its parts has, in the
way that the parity of three bits can be fixed while the bits are
pairwise independent~\cite{williams2010,rosas2019}.

Physics has a name for the sharpest version of such an object. Three
particles can be bound when no two of them are: the Borromean binding of
Efimov physics, with its tower of trimers at a two-body
resonance~\cite{efimov1970,nielsen2001,kraemer2006,kunitski2015}, of halo
nuclei, and, on lattices, of three distinguishable particles with on-site
attraction, where the band edge replaces the resonance and the triple
binds at an attraction below the pair threshold~\cite{mattis1984,mattis1986}.
In many-body physics the triple appears as the trion of three-flavour
attractive fermions, the cold-atom analogue of the baryon, with a
Cooper-problem analogue, the Cooper
triple~\cite{niemann2012,akagami2021,tajima2022}; as the charge-$6e$
condensate of three bound Cooper pairs~\cite{ge2024}; as the light
clusters of warm nuclear matter, where Pauli blocking dissolves the
deuteron before the triton~\cite{typel2010}; and, as the ordered-phase
cousin of Borromean binding rather than a realisation of it, as the
Borromean super-counterfluid of three-component mixtures, in which a
cubic coupling locks three inter-component phases at once so that the
three-component counterflow is superfluid while no component and no
pair is~\cite{blomquist2021,babaev2024,kuklov2026}.
A bound triple with no bound pair is a $3$-hyperedge with none of its
$2$-faces, a hypergraph that is not a simplicial complex, since a
simplex would carry its edges. It is also the strongest form the notion
of an emergent group interaction can take: not an effective coupling
that appears in a coarse-grained description, but a persistent composite
entity, a bound state, made by pairwise couplings alone.

This paper puts the mechanism on the graphs network science uses and
asks the question that decides whether an emergent group interaction
matters at large scale: does it survive the heterogeneity of a real
graph, or does disorder pin it to a few sites? A sparse random graph is
locally a tree, its local physics in the thermodynamic limit is that of
the Bethe lattice of the same degree, where the cavity method is exact~\cite{abouchacra1973,vazquez2023pre,chybook}, and on that lattice the attractive Hubbard model has been solved by dynamical mean-field
theory~\cite{keller2001,garg2005,toschi2005}, with
disorder~\cite{ioffe2010,feigelman2010} and with three
flavours~\cite{inaba2009,inaba2011}, where the triple appears as the
trion that competes with the colour superfluid of
pairs~\cite{rapp2007,inaba2009,inaba2011}. The Bethe lattice is also
where Anderson localisation has its exact solution, the cavity
recursion for the resolvent of Abou-Chacra, Anderson and
Thouless~\cite{abouchacra1973}, and that recursion applies to a
composite as it does to a particle, once the composite's hopping and its
sensitivity to the site energies are known. The paper builds those two
numbers from the few-body wavefunctions and puts the emergent hyperedge
through the transition. Section~\ref{sec:def} defines an emergent
hyperedge and its operational test. Sections~\ref{sec:model}
to~\ref{sec:l2} compute the pair and trimer thresholds on the Bethe
lattice, the pair ones in closed form, in the two sectors that an
expander has and a lattice lacks. Section~\ref{sec:rrg} runs the test on
finite random regular graphs, Sec.~\ref{sec:density} shows that quantum
statistics selects which sector a many-body phase sees,
Sec.~\ref{sec:fermions} that statistics can also forbid the object, and
Secs.~\ref{sec:site} and~\ref{sec:disorder} where it nucleates on a
heterogeneous graph. Section~\ref{sec:loc}, the anchor of the paper, shows that the emergent
hyperedges localise before their parts.

It helps to say at the outset what here is a new instance of known
physics, what is new as a result of calculation, and what is new. Known,
and recovered on a new graph, are four things: that a band edge alone
produces Borromean binding of three distinguishable particles, the result
of Mattis and Rudin on the cubic lattice~\cite{mattis1984}; that two
identical fermions and a third particle do not bind, Mattis's
no-go~\cite{mattis1986}, which the exchange argument of
Sec.~\ref{sec:fermions} shows to hold on any graph; that a lattice has no
Efimov tower, which was Mattis's point and which the count of Sec.~\ref{sec:uniform} confirms; and that the three-flavour gas has a
trionic phase at strong coupling~\cite{rapp2007,inaba2011}, here extended
to zero density with a point-trimer estimate. New as results of
calculation are the closed-form pair thresholds in both sectors,
Eqs.~\eqref{eq:U2} and~\eqref{eq:U2L2}, their exact large-$K$ limits from
the semicircle density, the monotonic growth of the Borromean window with
the degree, the thresholds at a single deep site and on a disordered
tree, the direct verification on random regular graphs conditioned on
girth, and an exact cavity for the percolation of the emergent
hyperedges. New are three things. For network science, the definition of
an emergent hyperedge with a test in terms of a three-point statistic,
and the demonstration that the sparse random graphs of the field produce
one from the simplest pairwise dynamics. For physics, that on an expander
a composite has two thresholds, separated by a spectral gap that Friedman's theorem keeps open on random
regular graphs and that the Bethe lattice saturates, by the Alon--Boppana
bound~\cite{friedman2008,alon1986,nilli1991}, and that
quantum statistics selects which one a many-body phase sees, a bosonic
composite condensing into the Perron mode and binding at the uniform
threshold, a fermionic one filling a band and binding at the
square-summable one. And, for both, the localisation transition of the
emergent hyperedge: a composite made by pairwise couplings localises at a
disorder an order of magnitude below the one that localises its
constituents, so that the liquid of emergent hyperedges is a glass over a
range of heterogeneity in which the pairwise network that produced them
still conducts.

\section{What an emergent hyperedge is}
\label{sec:def}
Let $G$ be a graph and $H$ a Hamiltonian whose terms live on the vertices
and edges of $G$ only, $H=\sum_{i}h_{i}+\sum_{\langle ij\rangle}h_{ij}$,
acting on $k$ particles. Write $E_{k}$ for the $k$-body ground energy and
$e_{\min}$ for the one-body one. For every partition of the $k$ particles
into clusters the sum of the clusters' ground energies is an energy the
$k$ particles can reach without binding as a whole. A $k$-body bound
state is \emph{irreducible} when $E_{k}$ lies below every such sum with
more than one cluster, and it is an \emph{emergent hyperedge without
faces}, the Borromean case, when in addition no proper sub-cluster is
bound at all, $E_{j}=j\,e_{\min}$ for $1<j<k$. With a coupling $U$ that
tunes the attraction, each cluster size has a threshold $U_{k}$ at which
$E_{k}$ first drops below $k\,e_{\min}$, and the hyperedge without faces
exists in the window $U_{k}<U<U_{k-1}$, whose width measures how
strongly the graph favours the group over its parts.

On a finite graph of $N$ vertices the definition has an operational
form. Let $P_{k}$ be the probability, in the $k$-body ground state, that
all $k$ particles occupy one vertex. A bound composite is a localised
relative motion carried by a centre of mass that spreads over the graph,
so $P_{k}$ stays finite as $N$ grows; an unbound one spreads its
particles independently and $P_{k}$ vanishes, as $N^{1-k}$ for $k$ free
particles and as $1/N$ for a bound $(k-1)$-cluster and a free particle. An emergent hyperedge is a
$k$-fold coincidence that survives the thermodynamic limit while every
$(k-1)$-fold one dies. This is the analogue, for a ground state, of the
information-theoretic notion of an irreducible dependence, a synergy that
no subset of the variables carries~\cite{williams2010,rosas2019}, with
one difference: the analogy is at the level of ground states, not of one
state's correlations. The $(k-1)$-body ground states are free while the
$k$-body one is bound; inside the bound $k$-body state the $(k-1)$-fold
coincidences are of order one, as they must be for a bound cluster, and
it is the $(k-1)$-body problems, not the $(k-1)$-point functions of the
$k$-body state, that carry no trace of the object.

The point of the definition is that a group interaction on a network
need not be part of the system's rules: it can be made by pairwise
dynamics on the underlying graph, and what is then observed is a
$k$-fold coincidence that survives the thermodynamic limit when the
$(k-1)$-fold ones do not. Whether classical pairwise dynamics do the
same, with a three-body activation that percolates when no pairwise
process does or the parity dependences of Boolean dynamics as
candidates, is left open here; the definition applies to them unchanged
once the ground state is replaced by the stationary state. The rest of
the paper is the case $k=3$ on the sparse random graphs of network
science, with the simplest pairwise $H$: hopping along the edges and
attraction on the vertices.

\section{Model and the two thresholds}
\label{sec:model}
\begin{figure}[t]
\includegraphics[width=\columnwidth]{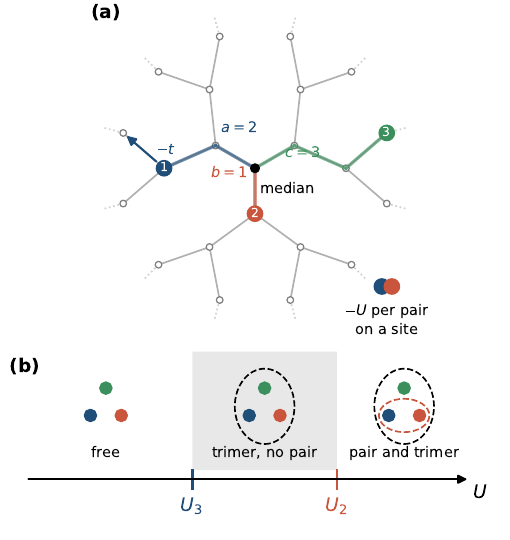}
\caption{The problem. (a) Three distinguishable particles on the Bethe
lattice of degree $3$ ($K=2$), hopping with amplitude $-t$ and attracting
with $-U$ per pair on a site. The median vertex is the unique vertex on
the three geodesics, and the distances $(a,b,c)$ from it are the
coordinates of the uniform sector, Sec.~\ref{sec:uniform}. (b) The
question is the order of the two thresholds: the attraction $\Ut$ at which
three particles first bind and $\Up$ at which two do. Between them
(shaded) the trimer exists and no pair does, the Borromean window; on a
tree each threshold is asked in two sectors, Sec.~\ref{sec:model}.}
\label{fig:scheme}
\end{figure}

The graph is the Bethe lattice of
degree $z=K+1$, the infinite tree in which every vertex has $K+1$
neighbours, with $K$ the branching. Distinguishable particles
$p=1,\dots,m$ hop between neighbouring vertices with amplitude $-t$ and
attract on site,
\begin{equation}
  H=-t\!\sum_{\langle ij\rangle,p}\!
  \left(c^{\dagger}_{ip}c_{jp}+\text{h.c.}\right)
  -U\sum_{i,\,p<q}n_{ip}n_{iq}
  +\sum_{i}\eps_{i}n_{i},
  \label{eq:H}
\end{equation}
where h.c.\ stands for the Hermitian conjugate of the preceding term,
\begin{equation}
  n_{ip}=c^{\dagger}_{ip}c_{ip},\qquad n_{i}=\sum_{p}n_{ip},
  \label{eq:n}
\end{equation}
and $\eps_{i}$ is a site energy, zero except where stated. Figure~\ref{fig:scheme}
shows the setting and the question. For three flavours of
fermions the flavours are the labels $p$ and the statistics plays no role;
I call the three-body bound state a trimer throughout and reserve trion for
the many-body literature. The single-particle spectrum of the tree is the
band $|E|\le2\sqrt{K}\,t$ with the Kesten--McKay density of
states~\cite{kesten1959,mckay1981}, so $m$ free particles have their
continuum edge at $-2m\sqrt{K}\,t$. The constant vector is an eigenvector
of the hopping with energy $-(K+1)t$, below the band, and it is not
square-summable: it is the Perron mode of a finite random regular graph,
where it is a legitimate state. A composite of $m$ particles is bound when
its energy lies below $-2m\sqrt{K}\,t$, and on a tree that can be asked in
two sectors. In the \emph{square-summable} sector the composite is an
ordinary normalisable state, its centre of mass at the band bottom
$-2\sqrt{K}\,t_{\rm eff}$ of its own effective hopping. In the
\emph{uniform} sector the relative motion is normalisable but the centre of
mass is the Perron mode, at $-(K+1)t_{\rm eff}$, lower. The uniform
threshold is therefore the lower of the two. The gap between the Perron
energy $-(K+1)t$ and the band edge $-2\sqrt{K}\,t$ is the spectral gap of a non-amenable graph. On random regular graphs
it stays open in the thermodynamic limit, by Friedman's theorem on
Alon's conjecture~\cite{friedman2008}, and the Bethe lattice saturates
it: the Alon--Boppana bound says that the bulk of any $(K+1)$-regular
family reaches $2\sqrt{K}\,t$, so no graph of that degree has a larger
gap~\cite{alon1986,nilli1991}. On a lattice the uniform mode is the $P=0$
Bloch state, inside the band, and the two thresholds coincide. Both are physical, and Sec.~\ref{sec:density}
shows which one a many-body phase sees: a condensate of bosonic
composites occupies the Perron mode and binds at the uniform threshold,
a Fermi liquid of composites fills a band and binds at the
square-summable one. Every number below says which sector it belongs to.
I set $t=1$.

\section{Uniform sector}
\label{sec:uniform}
\begin{figure}[t]
\includegraphics[width=\columnwidth]{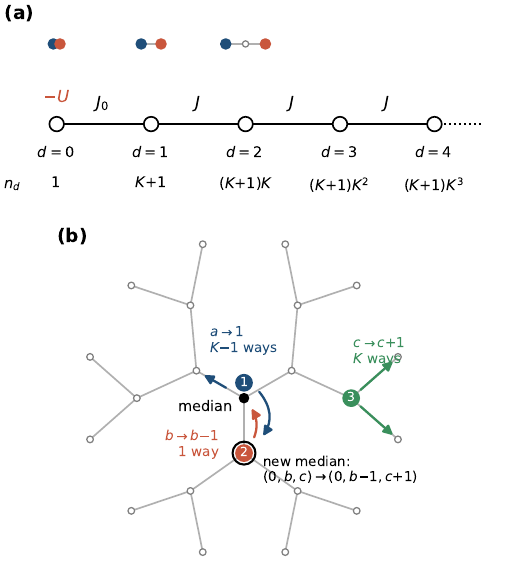}
\caption{The uniform-sector calculation. (a) The pair problem reduced to
the half line of the distance $d$: site energy $-U$ at $d=0$, hopping
$J_{0}=2\sqrt{K+1}\,t$ on the first link and $J=2\sqrt{K}\,t$ beyond, with
the orbit sizes $n_{d}$ and, above the first three sites, the
configurations they stand for. (b) The moves in the three-body orbit
$(a,b,c)$, for $K=2$: a particle steps away from the median with $K$
choices or towards it with one; a particle at the median enters a fresh
branch with $K-1$ choices; and a particle at the median stepping towards
another relocates the median (dashed ring), changing two distances at
once. The hop counts $M_{ss'}$ are these multiplicities, and the orbit
sizes $n(a,b,c)$ balance them.}
\label{fig:calc}
\end{figure}

The tree has a transitive automorphism group,
and the uniform sector is the set of wavefunctions constant on its orbits,
the analogue of zero total momentum. For two particles the orbit is their
distance $d$, with $n_{d}=(K+1)K^{d-1}$ configurations per particle at
$d\ge1$. For three particles the orbit is the triple $(a,b,c)$ of distances
from their median, the unique vertex on all three geodesics, with the
particles at positive distance in distinct branches; the orbit has
$n(a,b,c)=(K+1)^{(k)}K^{a+b+c-k}$ configurations per median, where $k$ is
the number of positive entries and $x^{(k)}$ the falling factorial. The
Hamiltonian acts on orbit functions $f$ through a matrix $M$ of hop
counts, Fig.~\ref{fig:calc}(b), that satisfies detailed balance with
respect to $n$, so $g=\sqrt{n}\,f$
obeys a symmetric problem with kinetic elements $-t\sqrt{M_{ss'}M_{s's}}$.
The two-body sector is a half line in $d$, Fig.~\ref{fig:calc}(a), with
hopping $J=2\sqrt{K}\,t$ for $d\ge1$, hopping $J_{0}=2\sqrt{K+1}\,t$ on the
first link and site energy $-U$ at $d=0$; the cavity element of the half line at the band edge is
$1/J$, and the bound-state condition $-U-E-J_{0}^{2}/J=0$ at $E=-2J$ gives
\begin{equation}
  \Up^{\rm u}=2J-\frac{J_{0}^{2}}{J}=\frac{2(K-1)}{\sqrt{K}}\,t .
  \label{eq:U2}
\end{equation}

It vanishes on the line, $K=1$, as it must, and at large $U$ the pair
energy is $-U-4(K+1)t^{2}/U$, the on-site shift $-2(K+1)t^{2}/U$ plus the
Perron energy of a pair hopping with amplitude $2t^{2}/U$, which is the
sector speaking. The three-body sector I diagonalise with the orbits
truncated at $a+b+c\le L$. The truncated continuum lies above the true
edge, so the smallest $U$ at which the ground energy drops below
$-6\sqrt{K}\,t$ is an upper bound on $\Ut^{\rm u}$ that decreases with $L$;
it is converged at $L=60$ and unchanged at $120$ and $180$. The reduced
model reproduces exact diagonalisation on a ring at $K=1$ to eight digits,
Eq.~\eqref{eq:U2}, and the large-$U$ expansion
$E_{3}=-3U-3(K+1)t^{2}/(2U)+O(t^{4}/U^{3})$. The same sector answers
the Efimov question. The effect is a lattice effect in the sense of
Mattis~\cite{mattis1986}, a consequence of the band edge rather than of
a resonance, and a tower is not expected on a graph; the count confirms
it. The states below the three-body edge at $U=\Up^{\rm u}$, where the
pair is at resonance, number one for $K=2$ and $3$, unchanged from
$L=60$ to $180$, with the next state approaching the edge as $L^{-2}$, a
continuum state. The square-summable sector cannot be used for this
count, since the centre-of-mass band of the bound trimer fills the
region below the edge.

\section{Square-summable sector}
\label{sec:l2}
Fix a root vertex and take the
wavefunctions invariant under the automorphisms that fix it, which relabel
the children of every vertex independently. A configuration is reduced to
canonical form by relabelling branches in order of first use along the
particle paths, its orbit measure is a product of falling factorials over
the vertices of the tree spanned by the root and the particles, and the
symmetrisation is as before. Every state in this sector is square-summable
and the sector contains the radial part of every bound state. The
thresholds follow from the Birman--Schwinger principle: with $V$ the
diagonal operator that counts coincident pairs and $H_{0}$ the Hamiltonian
at $U=0$, a bound state below $E$ exists when $U\lambda_{\max}\ge1$, where
$\lambda_{\max}$ is the largest eigenvalue of
$V^{1/2}(H_{0}-E)^{-1}V^{1/2}$, so the threshold is $1/\lambda_{\max}$ at
the edge. For the pair the operator is the propagator between doubly
occupied sites, $b_{d}=\langle ii|(H_{0}+4\sqrt{K}\,t)^{-1}|jj\rangle$ at
distance $d$, a radial kernel and hence a function of the adjacency matrix
of the tree; its norm is its value on the spherical function at the band
edge, and its uniform-sector eigenvalue is its value on the constant
vector,
\begin{equation}
  \frac{1}{\Up}=\sum_{d}b_{d}\,P_{d}(2\sqrt{K}),\qquad
  \frac{1}{\Up^{\rm u}}=\sum_{d}b_{d}\,n_{d},
  \label{eq:U2L2}
\end{equation}
with $P_{d}(2\sqrt{K})=K^{d/2}[(K+1)+d(K-1)]/K$ for $d\ge1$ the distance
polynomial of the tree at the edge. The two thresholds are one kernel read
on two centre-of-mass functions. The $b_{d}$ are a double integral over
the Kesten--McKay density with the spherical functions, the first sum
converges geometrically and gives the pair column of Table~\ref{tab:thr} to
six digits, and the second reproduces Eq.~\eqref{eq:U2}. The trimer has no
such reduction and I compute $\lambda_{\max}$ on the truncated orbit space
at $L=30$ to $60$ and extrapolate with a fitted power of $L$, close to
$2$; the same extrapolation applied to the truncated pair problem
reproduces Eq.~\eqref{eq:U2L2} to $10^{-3}$, and the same routine
reproduces the bisection thresholds of the uniform sector to their grid.

\begin{table}[t]
\caption{Pair and trimer thresholds on the Bethe lattice of branching $K$
(degree $z=K+1$), in units of $t$, in the square-summable sector
($\Up$ from Eq.~\eqref{eq:U2L2}, $\Ut$ extrapolated in $L$,
uncertainty in the last digit) and in the uniform sector ($\Up^{\rm u}$
from Eq.~\eqref{eq:U2}, $\Ut^{\rm u}$ by bisection on a grid of
$2\times10^{-3}$). The last column is the cross-sector ratio of a
square-summable trimer to a uniform pair, the one that governs the
dilute three-flavour gas, Sec.~\ref{sec:density}.}
\label{tab:thr}
\begin{ruledtabular}
\begin{tabular}{rrrrrrrrr}
$K$ & $z$ & $\Up$ & $\Ut$ & $\Ut/\Up$ & $\Up^{\rm u}$ & $\Ut^{\rm u}$ & $\Ut^{\rm u}/\Up^{\rm u}$ & $\Ut/\Up^{\rm u}$ \\
\hline
 2 &  3 & 1.966 & 1.369 & 0.696 & 1.414 & 1.310 & 0.926 & 0.968 \\
 3 &  4 & 3.251 & 2.159 & 0.664 & 2.309 & 2.073 & 0.898 & 0.935 \\
 4 &  5 & 4.269 & 2.741 & 0.642 & 3.000 & 2.637 & 0.879 & 0.914 \\
 5 &  6 & 5.136 & 3.218 & 0.627 & 3.578 & 3.103 & 0.867 & 0.899 \\
 6 &  7 & 5.904 & 3.633 & 0.615 & 4.083 & 3.508 & 0.859 & 0.890 \\
10 & 11 & 8.402 & 4.949 & 0.589 & 5.692 & 4.800 & 0.843 & 0.869 \\
\end{tabular}
\end{ruledtabular}
\end{table}

\begin{figure}[t]
\includegraphics[width=\columnwidth]{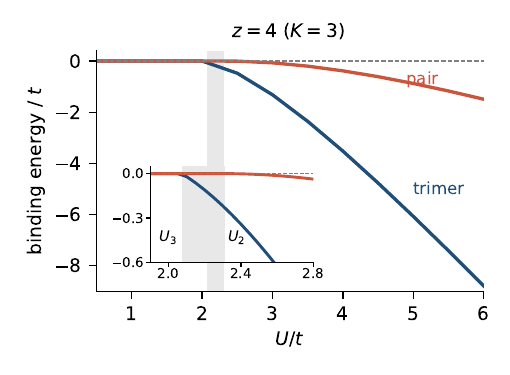}
\caption{Binding energies in the uniform sector of the Bethe lattice of
degree $z=4$ ($K=3$). Trimer energy measured from the three-particle
continuum edge, $E_{3}+6\sqrt{K}\,t$ (dark line), and pair energy measured
from the two-particle edge, $E_{2}+4\sqrt{K}\,t$ (light line), against the
on-site attraction $U$. The trimer binds at $\Ut^{\rm u}=2.07\,t$, the pair
at $\Up^{\rm u}=2.31\,t$, Eq.~\eqref{eq:U2}. Between the two, shaded, three
particles are bound and no two are.}
\label{fig:bind}
\end{figure}

Table~\ref{tab:thr} and Fig.~\ref{fig:bind} contain the main result. On
every tree with $K\ge2$ the trimer binds before the pair in all three
comparisons: square-summable trimer against square-summable pair, with a
window that widens from $30\%$ of $\Up$ at $K=2$ to $41\%$ at $K=10$;
uniform against uniform, from $7\%$ to $16\%$; and, the comparison that
Sec.~\ref{sec:density} shows to govern the dilute three-flavour gas, a
square-summable trimer against a uniform pair, from $3\%$ to $13\%$. The
first two are the tree's versions of the cubic-lattice result; the third
has no lattice counterpart, and it is the smallest, because the
condensed pair takes the Perron energy while the fermionic trimer does
not. Above $\Up$ the trimer stays below the continuum of a bound pair and
a free particle, $E_{2}(U)-2\sqrt{K}\,t$ in the uniform sector, where the
pair carries the Perron centre of mass and the third particle,
normalisable relative to it, enters at the band edge; so the trimer is
the ground composite for all $U>\Ut$. The large-$K$ limit is the
infinite-coordination limit of dynamical mean-field
theory~\cite{keller2001,garg2005}, with $t^{*}=\sqrt{K}\,t$ fixed and a
semicircular density of states. There the uniform pair threshold is
$\Up^{\rm u}\to2t^{*}$ from Eq.~\eqref{eq:U2}, and the square-summable one
keeps only the $d=0$ term of Eq.~\eqref{eq:U2L2}, since $b_{d}P_{d}$
vanishes as $K^{-d/2}$, so $1/\Up\to b_{0}$ with $b_{0}$ the double
integral of the semicircle at the two-body edge, $\Up\to3.308\,t^{*}$ and
$\Up/\Up^{\rm u}\to1.654$. The trimer ratios of Table~\ref{tab:thr} are
still falling at $K=10$; fits of the form $a+b/\sqrt{K}$ extrapolate them
to about $0.50$ in the square-summable sector, $0.78$ in the uniform one
and $0.79$ across sectors, but the same fit applied to the pair ratio,
whose limit is known, gives $1.54$ for $1.654$, so these are indications
that the infinite-coordination three-flavour gas keeps a finite Borromean
window, not values; the three-body problem on the semicircle is the
calculation that would fix them. On the cubic lattice the same Birman--Schwinger calculation in a
box of relative coordinates gives $\Up=7.914\,t$, the Watson integral, and $\Ut=5.158\,t$
(converged to the last digit at box radius $6$), a window of $35\%$, the
number behind the result of Mattis and Rudin~\cite{mattis1984}. The tree
of degree $6$ has a window of $37\%$ in the square-summable sector, close
to the cubic one, and of $13\%$ in the uniform sector, narrower
because that sector's pair threshold is the lower one. This is also
where experiment stands. The model is realised, with its three flavours,
by $^{6}$Li, whose three lowest hyperfine states have attractions tuned
together by a magnetic field, load into optical lattices, and show their
three-body physics in recombination
losses~\cite{ottenstein2008,huckans2009,williams2009}, and a speckle or
quasi-periodic potential adds site disorder~\cite{billy2008,roati2008}.
An optical lattice is cubic, so what such an experiment tests is the
cubic-lattice window, $35\%$, and not the graph; the expander content,
the two sectors, needs a non-amenable graph, and its natural home is a
hyperbolic lattice in circuit quantum
electrodynamics~\cite{kollar2019,boettcher2020}, where the separation of
the uniform mode from the bulk band is already the observable and where
the composites, being photonic, are bosonic and see the uniform
thresholds.

\section{Finite random regular graphs}
\label{sec:rrg}
This section runs the test of Sec.~\ref{sec:def} on actual random graphs, for the square-summable sector.
On a finite random regular graph the Perron mode is a normalisable state,
three free particles have their ground state at $-3(K+1)t$, and in the
Borromean window the trimer is not the ground state of three particles but
a state bound relative to the bulk continuum, coupled to the
Perron-containing states with weight $O(1/N)$. To see it I diagonalise
three and two distinguishable particles on random $3$-regular graphs in
the sector orthogonal to the Perron mode on every particle, with a Lanczos
product that applies the projector, and use as diagnostic the probability
that all three particles sit on one site in the three-body ground state
and that both do in the two-body ground state. A bound composite keeps a
finite probability as $N$ grows; an unbound one loses it. Two features of
finite random regular graphs enter. Their lowest bulk states sit on rare
regions, short cycles first, which act as the deep sites of
Sec.~\ref{sec:site} and bind pairs below the clean threshold; I therefore
condition the graphs on girth at least $6$ by rejection sampling. And the
level spacing at the band edge closes only as $N^{-2/3}$, so the approach
to the thermodynamic limit is slow. With that, at $U=1.7\,t$, inside the
square-summable window $1.37<U/t<1.97$, the triple probability is $0.12$,
$0.13$ and $0.14$ at $N=100$, $150$ and $200$, three hundred times its
value at $U=0$, while the pair probability falls from $0.12$ to $0.10$,
$0.08$ and $0.08$ at $N=100$, $200$, $400$ and $800$, six times its
$U=0$ value at $N=100$ and still falling; at $U=2.3\,t$, above $\Up$, the
pair probability is $0.24$ at every $N$, and at $U=1.0\,t$, below both
thresholds, both probabilities fall with $N$. The trimer is bound and the
pair is not, on actual random regular graphs, in the sector the
square-summable thresholds describe, which is the test of
Sec.~\ref{sec:def} passed: a three-point coincidence that survives the
thermodynamic limit, produced by couplings that live on the edges of the
graph, while the two-point coincidences decay as those of free
particles. The uniform-sector thresholds are not tested by this
construction, which projects the Perron mode out of every particle;
their verification on a finite graph would need the Perron-containing
sector, where the trimer is not the ground state and couples to the
bulk at $O(1/N)$, a separate problem.

\section{Which composites a many-body phase forms}
\label{sec:density}
The two sectors acquire their meaning at finite density, and this section
is the evidence for the claim of the introduction that statistics selects
the sector. A condensate of bosonic composites occupies the Perron mode: the BCS mean field of two
colours with attraction $U$ on the Kesten--McKay density of states, solved
for the gap and the chemical potential $\mu$ at density $n$ per colour, has
$2\mu\to E_{2}^{\rm u}$, the uniform-sector pair energy, as $n\to0$ (to
five digits at $n=10^{-4}$), not the square-summable one. A Fermi liquid
of composites instead fills a band from its square-summable bottom, the
Perron mode being a single state. With three colours the composites are
pairs, which condense, and trimers, which are fermions, so the
zero-density ordering of Table~\ref{tab:thr} asks whether a Fermi liquid
of trimers or a colour superfluid, two colours paired and the third free,
has the lower energy at small $n$, the question that the many-body
literature settles at strong coupling and half
filling~\cite{rapp2007,inaba2011}.

I compare the two at the level of their composites' energies. The trimer is
a point fermion with hopping $t_{3}$ on the tree, defined from the two
sectors' energies at the same $U$ by
$E_{3}-E_{3}^{\rm u}=(K+1-2\sqrt{K})\,t_{3}$, with $E_{3}$ the
square-summable energy extrapolated in $L^{-2}$ from $L=30$ and $45$ and
$E_{3}^{\rm u}$ the uniform one at $L=60$; its band is Kesten--McKay with
bottom $E_{3}$ and its energy per site at density $n$ is that band filled
to $n$. The superfluid is the mean-field energy of the two paired colours
at density $n$ plus a free third colour at $n$. The Hartree energy
$-3Un^{2}$ is common to every uniform phase and is dropped from both. A
mixed phase with a fraction $x$ of the particles in trimers is allowed, and
the ground state is the minimum over $x$. The description holds while the trimers are dilute on the scale of
their size, a few lattice spacings near threshold: at $n=0.1$ per colour
and a trimer radius of $2$, whose ball holds $10$ sites at $K=2$, the
trimers overlap, so I use it for $n$ up to $0.05$ and quote $0.1$ only
as the point where it fails.

\begin{figure}[t]
\includegraphics[width=\columnwidth]{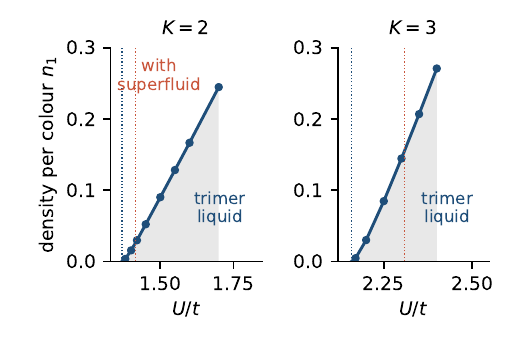}
\caption{Density $n_{1}$ per colour at which the colour superfluid enters,
against the attraction $U$, for $K=2$ and $3$. Below the line (shaded) a
Fermi liquid of trimers is the ground state; above it the colour
superfluid is a minority component, below $5\%$ of the particles up to
$n=0.3$. The line starts at $\Ut$ and crosses $n=0.05$, the limit of the
composite description, at $U=1.45\,t$ ($K=2$) and $2.22\,t$ ($K=3$);
at $n=0.1$ trimers of radius $2$ overlap.
Dotted lines mark $\Ut$, $\Up^{\rm u}$ and $\Up$ from Table~\ref{tab:thr}.}
\label{fig:phase}
\end{figure}

At zero density the trimer liquid wins for every $U>\Ut$. Between $\Ut$
and $\Up^{\rm u}$ no pair is bound in either sector and the superfluid
reduces to free fermions; above $\Up^{\rm u}$ the trimer lies below a
condensed pair and a free particle, $E_{3}<E_{2}^{\rm u}-2\sqrt{K}\,t$, by
a margin that grows with $U$. The superfluid enters at a density
$n_{1}(U)$ that vanishes at $\Ut$ and rises steeply, Fig.~\ref{fig:phase}:
for $K=2$, $n_{1}=0.003$ at $U=1.38\,t$, $0.015$ at $1.40\,t$, $0.05$
at $1.45\,t$ and $0.09$ at $1.50\,t$, so it crosses $0.05$ at
$U=1.45\,t$ and $0.1$ at $1.51\,t$; for $K=3$, $n_{1}=0.004$ at
$U=2.17\,t$, $0.03$ at $2.20\,t$ and $0.08$ at $2.25\,t$, crossing
$0.05$ at $2.22\,t$ and $0.1$ at $2.26\,t$. The onset is where
the trimer Fermi level reaches the binding energy,
$E_{3}+t_{3}\lambda_{F}(n_{1})=\min(-6\sqrt{K}\,t,\;E_{2}^{\rm u}-2\sqrt{K}\,t)$,
with $\lambda_{F}(n)$ the Fermi level of the Kesten--McKay band at filling
$n$. Above $n_{1}$ the superfluid is a minority component: the trimer band
is narrow, $t_{3}\approx0.17\,t$ at $K=2$ and $0.14\,t$ at $K=3$ near
threshold, so the trimer Fermi level rises slowly and the paired component
takes only the excess, below $3\%$ of the particles at $n=0.1$ and below
$5\%$ at $n=0.3$ for every $U>\Ut$ in the range. The trionic phase of the strong-coupling lattice
gas~\cite{rapp2007,inaba2011} therefore extends to zero density for all
$U>\Ut$, the colour superfluid enters from finite density, and the
zero-density boundary of that phase diagram is the cross-sector
comparison of Table~\ref{tab:thr}: between $\Ut$ and $\Up^{\rm u}$ the
trimer liquid faces no bound pair at all, and above $\Up^{\rm u}$ it
faces a condensed pair whose energy is the uniform-sector one. That is
the window a trion liquid enjoys on a tree, a few per cent at $K=2$ and
$13\%$ at $K=10$, and it is the statement the lattice cannot make. The estimate leaves out the interaction between trimers, the
pairing of the free colour with a trimer's constituents, and the
fluctuations of the mean field; a three-colour cavity calculation at
finite density would fix $n_{1}(U)$ beyond it.

\section{Statistics as a selection rule}
\label{sec:fermions}
Emergence depends on the three
particles being distinguishable. In the sector antisymmetric under the
exchange of particles 1 and 2, with attraction only between unlike
particles, the ground energy never drops below the pair-plus-particle
continuum for $U\le14t$ at $K=2$ and $K=3$, and the strong-coupling limit
says why. With the pair on site $i$ and the third fermion on a
neighbour $j$, the hop of the pair's unlike particle to $j$ leaves a pair
at $j$ and a free fermion at $i$, a degenerate configuration, so the
composite moves by exchange at first order in $t$; the fermionic sign
makes the amplitude $+t$, opposite to a hop, and Pauli blocking removes
the site $i$. The relative motion at leading order is the tree with one
vertex removed plus a positive potential $t$ on its neighbours, whose
spectrum lies above $-2\sqrt{K}\,t$, and the pair itself moves only at
order $t^{2}/U$: no bound state. Nothing in this uses the tree. By Cauchy
interlacing the spectrum of any vertex-deleted subgraph lies inside the
band of the graph, and a positive potential raises it, so the argument
recovers Mattis's no-go~\cite{mattis1986} on any graph; the tree adds
nothing to it and the section is a check, not a result. The same exchange with the bosonic sign
$-t$ is what binds trimers in the one-dimensional Bose--Hubbard
model~\cite{valiente2010}, and on the three-dimensional lattice the
equal-mass $2+1$ problem has no discrete spectrum below the continuum at
strong coupling, a trimer requiring a mass ratio above a
threshold~\cite{abdullaev2026}, as in free
space~\cite{kartavtsev2007}. The Borromean trimer of the tree is a
three-flavour object: quantum statistics is a selection rule on which
groups can emerge from pairwise couplings. For electrons the rule says
that negative-$U$ centres in amorphous
semiconductors~\cite{anderson1975,weaire1971}, which bind spin-paired
electrons, cannot by themselves produce Borromean three-electron
centres; a third flavour, orbital or valley, or a mass imbalance is
needed.

\section{One deep site}
\label{sec:site}
\begin{figure}[t]
\includegraphics[width=\columnwidth]{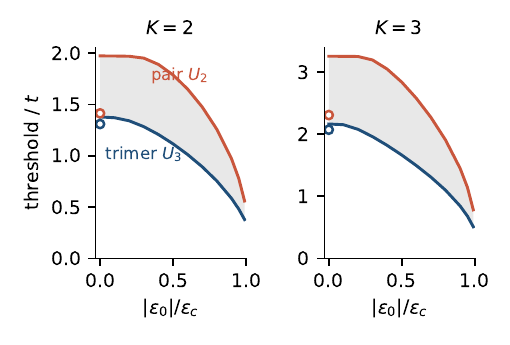}
\caption{Pair threshold $\Up$ (light line) and trimer threshold $\Ut$
(dark line) at a single site of energy $\eps_{0}$ on the Bethe lattice,
for $K=2$ and $K=3$, against the depth of the site in units of
$\eps_{c}=(K-1)t/\sqrt{K}$, the depth at which one particle binds.
Birman--Schwinger at the continuum edge, orbits truncated at $L=60$ for
the trimer and $120$ for the pair; the change from $L=40$ is below
$0.02\,t$ at $\eps_{0}=0$ and below $0.04\,t$ at $0.99\,\eps_{c}$, where
the states are least localised. Shaded: three particles bound, no two
bound. Open circles at $\eps_{0}=0$: the uniform-sector thresholds of
Table~\ref{tab:thr}.}
\label{fig:imp}
\end{figure}

Real graphs are heterogeneous, and heterogeneity enters at zero density
through its rare sites. Give the root the energy $\eps_{0}<0$: a static impurity is a
fourth particle that does not hop, the root-fixed sector is the natural
one, and its bound states are localised at the impurity. One particle
binds when $|\eps_{0}|>\eps_{c}=(K-1)t/\sqrt{K}$, from
$\eps_{0}-E-(K+1)t^{2}\mathcal{G}(E)=0$ with $\mathcal{G}$ the cavity
resolvent of the clean tree at the band edge, and the reduced model
reproduces this threshold, the impurity energy, and exact diagonalisation
on finite Cayley trees to five digits. For $-\eps_{c}<\eps_{0}<0$ the site
is too shallow to hold one particle, and Fig.~\ref{fig:imp} shows the
attraction it needs to hold two or three. Both thresholds fall from the
square-summable values of Table~\ref{tab:thr} at $\eps_{0}=0$ to zero at
$\eps_{0}=-\eps_{c}$, and the trimer threshold is the lower one at every
depth: $\Ut/\Up$ runs from $0.70$ at $\eps_{0}=0$ through a minimum of
$0.60$ near $0.8\,\eps_{c}$ and back to $0.68$ at $0.99\,\eps_{c}$ for
$K=2$, and from $0.66$ through $0.58$ to $0.65$ for $K=3$.

\section{Disordered tree}
\label{sec:disorder}
With box disorder $\eps_{i}\in[-W/2,W/2]$ on
every site, the thresholds of Fig.~\ref{fig:imp} classify the sites in
the isolated-site approximation of a Lifshitz tail, each site seen with a
clean environment. For $W=2\eps_{c}$, at $U=0.5\,\Up$ a fraction $0.14$
($K=2$) or $0.17$ ($K=3$) of the sites hold a trimer and no pair, against
$0.05$ or $0.07$ that hold a pair; at $U=0.7\,\Up$ the fractions are
$0.37$ and $0.13$, or $0.35$ and $0.15$; at $W=4\eps_{c}$ a quarter of the
sites hold one particle outright and the Borromean fraction at
$U=0.5\,\Up$ is $0.07$ or $0.09$. The approximation ignores the disorder
around a site, which shifts its thresholds in either direction, and it is
a statement about localised few-body states, not about the many-body
phases; those need the finite-density calculation. For a network the
lesson is where to look: emergent hyperedges nucleate on the rare deep
sites of a heterogeneous graph, and on a finite random graph, on its
short cycles, before they exist in the bulk.

Whether these localised trimers form a connected network is a
percolation problem in which the complexes are their supports, the balls
that hold $90\%$ of a trimer's one-body density, of radius $2$ for a deep
site well inside its window and $8$ or more for a shallow site. A site
within the radius of any Borromean centre is active, and the question is
whether the active sites percolate; on the tree this correlated site percolation is exact by a cavity
whose message carries the distance to the nearest centre inside a
subtree and the connection probability conditioned on the distance
outside, a dependent layer of the kind Ref.~\cite{chybook} uses for the
localisation landscape, and it agrees with Monte Carlo on random regular
graphs of $10^{5}$ vertices. With every centre of radius
$r$ the threshold falls as $K^{-r}$, by counting: $p_{c}=0.073$, $0.014$
and $0.003$ at $r=1$, $2$ and $3$ for $K=2$, and $0.024$, $0.0022$ and
$0.0002$ for $K=3$. The Borromean fractions above lie one to three orders
of magnitude higher, so the supports form a connected network wherever
the fraction is finite, for both $K$ and even when only the compact
trimers are admitted. This is a statement about geometry, not about a
phase: two adjacent deep sites are a six-body problem that the
isolated-site picture does not contain, and no observable is attached to
the connectivity beyond the absence of an isolated-trimer regime. The
question that carries an observable is whether the trimers move, and it
is the subject of the next section. For a network the lesson of this
section and the previous two is where to look: heterogeneity favours
emergence, a site too shallow to hold one particle holding three at
$0.6$ of the attraction at which it holds two and the short cycles of a
finite random graph playing the same role, so the first emergent
hyperedges of a network sit on its rare regions, a survey that averages
over the graph will not find them, and once they exist at a finite
fraction of the sites they are connected, with no isolated regime in
between.

\section{The emergent hyperedges localise before their parts}
\label{sec:loc}
Whether the emergent hyperedges move is a phase transition: the
Anderson localisation of a composite.
The transition itself is the single-particle one, so I state it first
in the form the cavity gives it. Put a particle on a site and ask
whether it leaks away. The resolvent recursion of the tree gives the
site a level, shifted by the real part of its self-energy, and a width,
the imaginary part, which is the rate at which the particle leaves into
the branches and does not return. In the extended phase that width stays
finite as the regulator $\eta$ goes to zero: the state is system-wide,
and on a finite random regular graph its inverse participation ratio
scales as $1/N$. In the localised phase the width vanishes with $\eta$:
the local spectrum is a set of poles, the particle stays within a
localisation length, and the inverse participation ratio stays finite.
Which phase holds is decided by the fate of an infinitesimal escape rate
injected at infinity. In the localised phase the imaginary parts are
infinitesimal and their recursion is linear: the imaginary part
transmitted along an edge is $t^{2}|G|^{2}$ times the sum of those
arriving from the $K$ branches beyond, so the injected rate is
multiplied along every path of length $n$ by a product of $n$ factors
$t^{2}|G|^{2}$ and summed over the $K^{n}$ paths. That sum is a
directed-polymer partition function on the tree with random weights,
broadly distributed because a site with $\eps_{i}-E-\operatorname{Re}\Sigma_{i}$
near zero is a resonance with a large $|G|$; the localised phase is its
frozen phase, in which the sum is carried by rare resonant paths and
still decays, and the transition is where the entropy of paths, $\ln K$
per step, first compensates the decay of the weights. Because of the
heavy tails the right measure of that decay is a fractional moment: the
localised phase is stable while the growth rate of the $\beta$-th moment
of the transmitted imaginary parts is below one for some $\beta$ in
$(0,1)$, the minimum sits at $\beta=1/2$ by a symmetry of the weight
distribution, and there the criterion reads $Kt\langle|G|\rangle=1$: the
transition is the disorder at which the mean modulus of the cavity Green
function per branch falls to $1/(Kt)$~\cite{abouchacra1973}. Approached
from the extended side the typical width vanishes with an essential
singularity, from the localised side the localisation length diverges as
$(W-W_{c})^{-1}$, and at fixed disorder below $W_{c}$ the same criterion
applied energy by energy gives the mobility edge, below which the states
of the Lifshitz tail, on rare regions of low site energy, are localised.
I evaluate the criterion by population dynamics of the resolvent, in
the implementation of Ref.~\cite{chybook}, which describes the
resolvent message and the mobility-edge scans, with the integrated
density of states from random regular graphs of $3000$ vertices built
with the same code.

A composite goes through the same transition with two numbers changed.
A trimer with hopping $t_{3}$ and one-body density $\rho$ sees a
site-energy shift $\sum_{k}\rho_{k}\eps_{k}$ of variance
$s_{3}^{2}W^{2}/12$, with $s_{3}^{2}=\sum_{d}\rho(d)^{2}/n_{d}$ between
$1.4$ near threshold and $3$ for a trimer on one site, so its Anderson
problem is the single-particle one on the same tree with hopping $t_{3}$
and disorder $s_{3}W$, exactly by scaling. Both numbers work against the
composite: its band is narrow, $2\sqrt{K}\,t_{3}$ against
$2\sqrt{K}\,t$, and its sensitivity to the site energies is up to
threefold, so a bare disorder that is a small fraction of the
single-particle bandwidth is several composite bandwidths.

The hopping $t_{3}$ of Sec.~\ref{sec:density} is defined from the two
sectors' energies, which assumes a rigid composite whose only difference
between the sectors is the energy of its centre of mass. That fails near
threshold, where the two sectors bind differently and the difference is
mostly uniform-sector binding energy: $t_{3}$ so defined is
non-monotonic, $0.16\,t$ at threshold, rising to $0.20\,t$ at $U=1.7t$
before falling, and for the pair between $\Up^{\rm u}$ and $\Up$ it is a
pure binding energy. Two checks fix the range in which it is a hopping.
At strong coupling the trimer moves by a third-order process,
$t_{3}=3t^{3}/(2U^{2})$, which gives $0.094\,t$ at $U=4t$ against
$0.087\,t$ from the sectors and $0.167\,t$ at $U=3t$ against $0.135\,t$
at $K=2$, the deviation being the next order. And where the whole trimer
band lies below the three-body continuum, the eigenvalues of the
root-fixed sector below the edge are the box-quantised band of the
centre of mass, whose width is $4\sqrt{K}\,t_{3}$ with no uniform-sector
energy involved. The band is a ladder of levels with the spacing of a
box, followed by a gap and one state that does not move with the
truncation, an internal excitation; the ladder's width at $L=40$ for
$K=2$ and $36$ for $K=3$, corrected for the finite number of levels,
gives $t_{3}=0.089\,t$, $0.149\,t$ and $0.201\,t$ at $U=4t$, $3t$ and
$2.5t$ for $K=2$, against $0.087\,t$, $0.135\,t$ and $0.166\,t$ from the
sectors, and $0.058\,t$, $0.089\,t$ and $0.114\,t$ at $U=5t$, $4t$ and
$3.5t$ for $K=3$, against $0.057\,t$, $0.083\,t$ and $0.102\,t$. The two
agree at strong coupling, the sector value falls $8\%$ below the band
width one bandwidth from the continuum and $20\%$ below it where the
band top approaches the continuum, and the deviation grows towards
threshold. The sector definition is therefore a hopping, to within
$20\%$, for $U\ge1.7t$ at $K=2$ and $U\ge2.5t$ at $K=3$, where the
trimer is compact, and I quote the localisation there with that
uncertainty, which enters $W_{c}^{(3)}$ and $W_{\rm glass}$ linearly
and leaves the comparison with the parts, an order of magnitude,
untouched. Nearer threshold a loose
composite is three nearly free particles seeing the bare disorder
rather than one particle seeing $s_{3}W$, and the rigid-composite
scaling is least valid exactly where the sector definition is least
reliable, so that region is left outside the description.

\begin{figure}[t]
\includegraphics[width=\columnwidth]{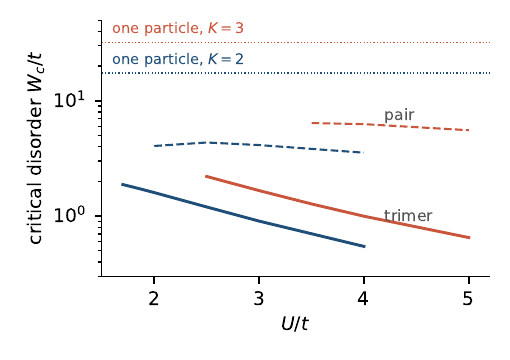}
\caption{Critical disorder $W_{c}$ above which every state of a
composite is localised, against $U$, for $K=2$ (dark) and $3$ (light):
trimer (full lines, drawn where the trimer is compact), pair where it is
bound in both sectors (dashed), single particle (dotted). The composite
values scale with the single-particle one, $W_{c}^{(m)}=W_{c}t_{m}/s_{m}$.}
\label{fig:wc}
\end{figure}

\begin{figure*}[t]
\includegraphics[width=\textwidth]{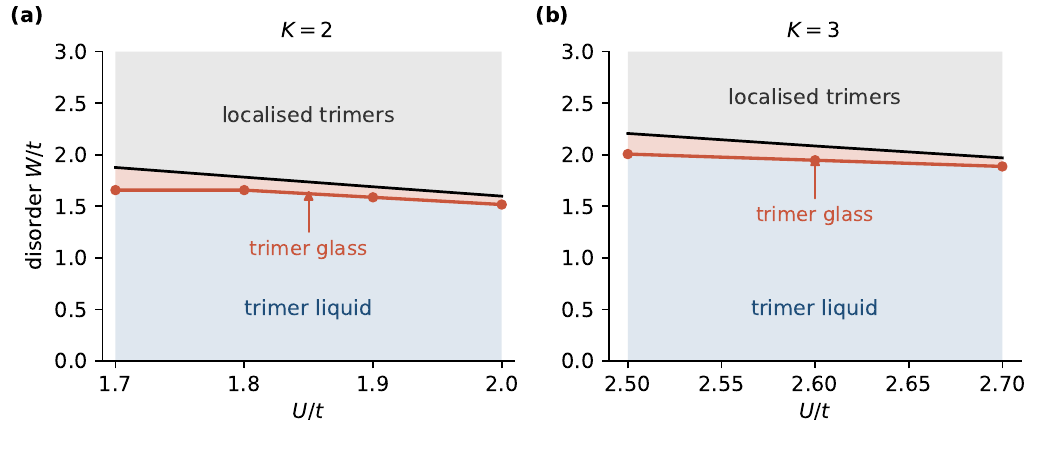}
\caption{Phase diagram of the Fermi liquid of trimers in the plane of
attraction $U$ and box disorder $W$, for (a) $K=2$ and (b) $K=3$, over the range
of $U$ in which the liquid exists below a finite density
$n_{1}(U)$. Trimer liquid: the occupied trimer states above the
Lifshitz tail are system-wide and the liquid conducts. Trimer glass: every occupied state of the liquid is
localised, $n_{1}(U)<n_{\rm tail}$, above the line $W_{\rm glass}(U)$
(circles). Localised trimers: above $W_{c}^{(3)}(U)$ (black line) no
trimer state is extended at any energy. A liquid of single particles at
the same density turns glassy only at $W=15t$ to $16t$ ($K=2$) and
$29t$ to $31t$ ($K=3$), above the frame. Dotted lines: the square-summable trimer threshold and the
uniform pair threshold. The single particle localises at $W=17.4\,t$
($K=2$) and $32\,t$ ($K=3$), off the scale.}
\label{fig:glass}
\end{figure*}

The disorder above which every trimer state is localised is
$W_{c}^{(3)}=W_{c}\,t_{3}/s_{3}$: at $K=2$ it is $1.9\,t$ at $U=1.7t$ and
$1.6\,t$ at $U=2t$, against $17.4\,t$ for a single particle and $4\,t$
for a pair once the pair is bound in both sectors; at $K=3$ it is
$2.2\,t$ at $U=2.5t$ and $1.7\,t$ at $U=3t$, against $32\,t$ and $6\,t$,
Fig.~\ref{fig:wc}. Below $W_{c}^{(3)}$ the localised trimer states are
those of the Lifshitz tail below the mobility edge, a fraction
$n_{\rm tail}$ of the trimer band that is below $0.5\%$ at $W=0.5\,t$
and between $3\%$ and $7\%$ at $W=t$ for $K=2$ and $1.7t\le U\le2t$. A
Fermi liquid of trimers at density $n$ is fully localised, a trimer
glass, when $n<n_{\rm tail}$; since the liquid exists only below
$n_{1}(U)$, Sec.~\ref{sec:density}, it is a glass at every density for
$W$ above the line $W_{\rm glass}(U)$ at which $n_{\rm tail}=n_{1}$,
which is $1.7\,t$ at $U=1.7t$ and $1.5\,t$ at $2t$ for $K=2$, and
$2.0\,t$ at $U=2.5t$ and $1.9\,t$ at $2.7t$ for $K=3$,
Fig.~\ref{fig:glass}. The comparison with the parts has to be made at
the same density, since comparing full-localisation disorders is close
to a statement about bandwidths, a narrow band localising first by
construction. The same population dynamics gives the glass line of a
Fermi liquid of single particles at the density $n_{1}(U)$, the disorder
at which its Fermi level falls below the mobility edge: $15\,t$ to
$16\,t$ for $K=2$ and $29\,t$ to $31\,t$ for $K=3$, the grey line of
Fig.~\ref{fig:glass}, an order of magnitude above the trimer's glass
line at every $U$. The phase diagram of the emergent hyperedges in the
$(U,W)$ plane therefore has three regions: a liquid, in which the
trimers above the tail are system-wide and the liquid conducts; a trimer
glass, in which every occupied state of the liquid is localised while
the trimer band still has extended states above the Fermi level; and,
above $W_{c}^{(3)}$, no system-wide trimer state at all. All three lie
at disorders where a liquid of single particles at the same density is
deep in its extended phase. Disorder therefore favours the trimer side
of the phase diagram twice: it nucleates the emergent hyperedges on its
deep sites, and it localises the liquid they form at a disorder an order
of magnitude below the one that localises single particles at the same
density, with the pair in between. Two caveats. The population dynamics
gives $W_{c}=17.4\,t$ at $K=2$, where large-scale studies of random
regular graphs give $18.2\,t$; the composite critical disorders scale
with it and carry the same $4\%$. And the effective disorder matches the
variance of the composite's site-energy shift, which is exact to first
order in $W$ and treats the composite as rigid; a trimer that deforms in
the disorder is more sensitive, not less, so the composite critical disorders are upper bounds in that respect.
For a network the conclusion is that an emergent group interaction is
more fragile to heterogeneity than the pairwise network that makes it:
the composite's band is narrower and its sensitivity to site energies
larger than its constituents', so the liquid of emergent hyperedges is a
glass at a disorder where single particles at the same density still
move freely, and a higher-order structure produced by dynamics should be
expected to be pinned, not transported, on any graph with moderate
heterogeneity.

\section{Conclusions}
\label{sec:conclusions}
A pairwise Hamiltonian on a sparse random graph produces a bound triple
with no bound pair, an emergent hyperedge without faces, by the band-edge
mechanism of Mattis and Rudin~\cite{mattis1984}; the pair thresholds are
exact, the trimer binds first for every degree, and the window widens
with the degree.

Random graphs are expanders, so a composite on them has
two thresholds, and quantum statistics selects which one a many-body
phase sees: pairs condense into the Perron mode and bind at the uniform
threshold, trimers fill a band and bind at the square-summable one, and
the cross-sector window, a few per cent at $K=2$ growing to $13\%$ at
$K=10$, is the one a Fermi liquid of trimers has against a colour
superfluid.

On a disordered graph the emergent hyperedges nucleate on
rare deep sites and localise before their parts: the compact trimer
undergoes the Anderson transition, the cavity transition of Abou-Chacra,
Anderson and Thouless~\cite{abouchacra1973} with the composite's hopping
and effective disorder, at $1.6\,t$ against $17\,t$ for a single
particle, and the liquid of trimers is a glass above a line near $1.5\,t$
where a liquid of single particles at the same density needs $15\,t$ to
$16\,t$.

A group interaction made by pairwise dynamics is an order of
magnitude more fragile to heterogeneity than the network that makes it.

\begin{acknowledgments}
All code that produced the numbers in this paper is available at
Ref.~\cite{code}. The calculations were assisted by Claude, an AI
assistant developed by Anthropic. Nodes \& Links Ltd provided support in
the form of salary for Alexei Vazquez but did not have any additional role
in the conceptualization of the study, the analysis, the decision to
publish or the preparation of the manuscript.
\end{acknowledgments}

\bibliographystyle{apsrev4-2}
\bibliography{refs}

\end{document}